\documentclass{l4dc2026}
\usepackage{booktabs}
\usepackage{multirow}
\usepackage{wrapfig}

\title{Zero-Shot Function Encoder-Based Differentiable Predictive Control }
\usepackage{times}
\usepackage{cleveref}

\coltauthor{%
\Name{Hassan Iqbal} \Email{hassan.iqbal@utexas.edu}\\
\Name{Xingjian Li} \Email{xingjian.li@austin.utexas.edu}\\
 \Name{Tyler Ingebrand} \Email{tyleringebrand@utexas.edu} \\
 \Name{Adam Thorpe} \Email{adam.thorpe@austin.utexas.edu} \\
 \Name{Krishna Kumar} \Email{krishnak@utexas.edu}\\
 \Name{Ufuk Topcu} \Email{utopcu@utexas.edu}\\
 \addr The University of Texas at Austin, 78712, TX, USA
\AND
\Name{J\'an Drgo\v na} \Email{jdrgona1@jh.edu}\\
\addr Johns Hopkins University, Baltimore, 21218, MD, USA}

\begin{document}

\maketitle

\begin{abstract}%
We introduce a differentiable framework for zero-shot adaptive control over parametric families of nonlinear dynamical systems. Our approach integrates a function encoder-based neural ODE (FE-NODE) for modeling system dynamics with differentiable predictive control (DPC) for offline self-supervised learning of explicit control policies. The FE-NODE captures nonlinear behaviors in state transitions and enables zero-shot adaptation to new systems without retraining. The DPC  efficiently learns control policies across system parameterizations, thus eliminating costly online optimization common in classical model predictive control. We demonstrate the efficiency, accuracy, and online adaptability of the proposed method across a range of nonlinear systems with varying parametric scenarios, highlighting its potential as a general-purpose tool for fast zero-shot adaptive control.
\end{abstract}


\section{Introduction}

Learning-based control methods rely on fixed models or predefined datasets of observations to train, which limits their ability to generalize when system parameters or environmental conditions change online. 
Systems such as aircraft flying through variable wind fields~\citep{beliaev2023development} or delivery drones carrying shifting payloads~\citep{palunko2011adaptive} require controllers that can adjust to new dynamics in real-time, without retraining or reoptimization. However, real-time adaptation remains challenging because most optimization-based and learning-based controllers depend on expensive, repeated online updates or gradient-based policy refinement. These computations introduce significant computational overhead, making them unsuitable for fast or safety-critical operation. Achieving zero-shot control, in which a learned policy adapts to new dynamics without additional online optimization, requires closed-loop control policies that can adjust instantaneously to variations in the underlying system dynamics.

We present a method for zero-shot control that combines function encoders (FE) with differentiable predictive control (DPC) to achieve adaptive closed-loop policies without costly online optimization. 
Parametric differentiable predictive control provides a gradient-based approach to optimal control via end-to-end learning of feedback control policies through backpropagation. However, its viability hinges on the underlying parametric dynamical system being fully known and differentiable, a condition often unmet in real-world problems due to noise, system degradation, and other uncertainties~\citep{le2024contact,list2025differentiability}.
To overcome this limitation, we condition the learned DPC control policy on a compact learned representation of the underlying system using function encoders \citep{ingebrand2024zero}. 
Function encoders represent the dynamics as a linear combination of learned neural ODE basis functions, thereby enabling zero-shot modeling by efficiently computing the corresponding basis coefficients online from limited system measurements. Conditioning the DPC policy on the function encoder coefficients allows the controller to adjust instantaneously to new and even unseen dynamics at runtime.

\paragraph{Related Work.}

Model predictive control (MPC)  remains the industry standard for controlling multi-input, multi-output constrained systems~\citep{Borrelli2017,rawlings2020model,samad2020industry,schwenzer2021review}. 
MPC excels in accuracy and reliability for a wide range of problems through receding-horizon optimization, but as control systems become increasingly complex, solving the associated online optimization problems repeatedly with accurate models can become prohibitively expensive. This has inspired work that focuses on hardware acceleration and algorithm optimization, such as~\citep{houska2011acado,Verschueren2021,adabag2024mpcgpu,zanelli2020forces,frison2020hpipm,wu2023simple}.
Requiring an accurate and fully differentiable dynamics model is often another bottleneck for MPC solvers. Integrating system identification or uncertainty quantification with classic MPC solvers~\citep{koller2018learning,fasel2021sindy,martinsen2020combining,ALSEYAB2008568} can serve as a solution; however, it also introduces additional cost and error accumulation, limiting practicality.

While accurate, the high online cost of MPC solvers limits their use in many real-time control applications. Learning-based and data-driven approaches~\citep{hewing2020learning,jiang2020learning,Berberich2021} have emerged as an effective alternative. These methods amortize control computation through offline training, enabling real-time deployment. Existing approaches can be broadly categorized into supervised learning-based approximate MPC, such as~\citep{hertneck2018learning,chen2018approximating,karg2020efficient,Pin01052013,li2025napi, hose2024parameter}, which learn policies directly from trajectories or expert demonstrations, and self-supervised methods~\citep{jin2020pontryagin,drgovna2022differentiable,onken2022neural,drgovna2024learning}, which rely on known differentiable dynamics and objectives for policy optimization.

Function encoders~\citep{ingebrand2025function,low2025function} provide a structured and principled framework for neural network-based function space approximation, with prominent applications in reinforcement learning~\citep{ingebrand2024zeroRL} and control~\citep{ward2025zero,li2025zero}. In this work, we focus on system identification~\citep{ingebrand2024zero}, which extends from prior classical~\citep{isermann2011identification,pintelon2012system} and learning-based methods such as NODE~\citep{chen2018neural,rahman2022neural}, SINDy~\citep{brunton2016discovering}, Koopman operator~\citep{korda2018convergence,klus2020data} and more. Function encoders offer retraining-free adaptability to different dynamics, an ability that existing methods often lack.

\begin{figure}[ht!]
    \centering
    \includegraphics[width=0.99\linewidth]{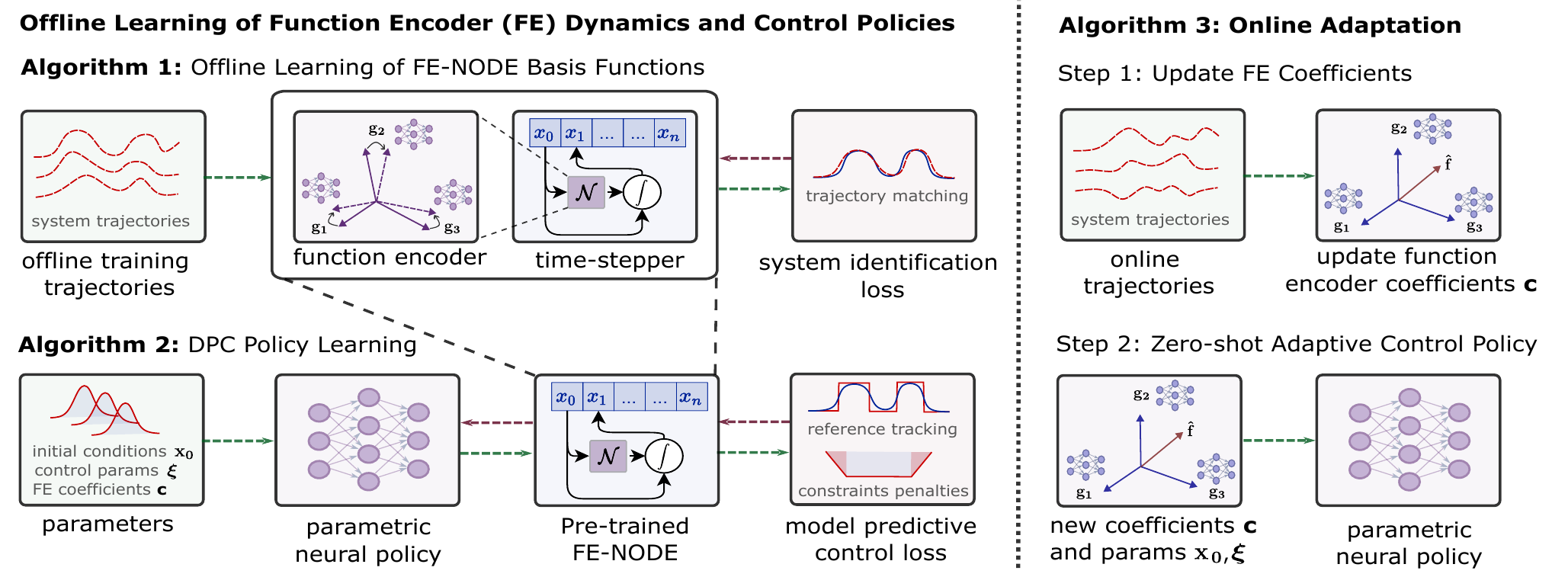}
    \vspace{-0.1cm}
    \caption{Conceptual diagram of the proposed Function-Encoder Differentiable Predictive Control. }
    \label{fig:FE-DPC_diagram}
    \vspace{-2em}
\end{figure}

\paragraph{Contributions.}
We propose a framework for zero-shot adaptive control that combines function encoder-based system identification with differentiable predictive control, as illustrated in~\Cref{fig:FE-DPC_diagram}. Our main contributions are:
1) \textit{Zero-shot control via function-encoder representation}.
We develop a differentiable control architecture in which system dynamics are represented using a compact FE-NODE basis and identified online through low-cost coefficient inference. A parametric DPC policy is trained offline over this learned dynamics space, enabling instantaneous online adaptation to previously unseen dynamics without retraining or reoptimization.
2) \textit{Scalable performance across nonlinear and complex benchmarks.}
We demonstrate the proposed method on a range of nonlinear control tasks-including the Van der Pol oscillator, two-tank level regulation, a stiff glycolytic oscillator, and a 12D quadrotor model. Across these settings, the learned policies maintain closed-loop stability under abrupt dynamics changes and achieve orders-of-magnitude faster inference compared to online MPC, while retaining competitive control performance.
3) \textit{Open-source implementation.}
We release all code, models, and training pipelines to enable reproducibility at \href{https://github.com/hassaniqbal209/DPCFunctionEncoder.git}{https://github.com/hassaniqbal209/DPCFunctionEncoder.git}

\section{Problem Formulation}

We consider a general class of parametric optimal control problems (pOCP) that take the  following continuous-time form

\begin{subequations}
\label{eq:pocp}
    \begin{align}
 \min_{\pi \in\Pi}  \ \  & \mathbb{E}_{ {\mathbf{x}}_0 \sim P_{\mathbf{x}_0}, \boldsymbol \xi \sim P_{\boldsymbol \xi}, \boldsymbol{\nu} \sim P_{\boldsymbol \nu} }\,
 \bigg( \int_{0}^{T} \ell(\mathbf{x}(t), \mathbf{u}(t); \boldsymbol{\xi}) \text{d} t + p_T (\mathbf{x}(T)) \bigg)
 \label{eq:pocp:objective} \\ 
 \text{s.t.} \ \ & \frac{\text{d} \mathbf{x}(t)}{\text{d} t} = {\mathbf{f}}( {\mathbf{x}(t)}, \pi({\mathbf{x}(t)}; \boldsymbol{\xi}, \boldsymbol{\nu}); \boldsymbol{\nu}),  
 \label{eq:pocp:x} \\
 \ & h({\mathbf{x}}(t); \boldsymbol{\xi}) \le 0, \quad
  g({\mathbf{u}}(t); \boldsymbol{\xi}) \le 0, \label{eq:pocp:g}
\end{align}
\end{subequations}

where $\boldsymbol{\xi}$ is a set of parameters that determine constraints and objective functions for the given problem,
$T$ is the prediction horizon, 
$h$ and $g$ are continuously differentiable state and control constraints, and the functions $\ell$ and $p_N$ define running and terminal costs. 
We denote $\mathbf{u}(t) = \pi(\mathbf{x}(t); \boldsymbol{\xi}, \boldsymbol{\nu})$ the control policy we aim to recover. 
Notably, we assume the distributions of the initial state $\mathbf{x}(0) = \mathbf{x}_0$ and problem parameters $\mathbf{x}_0 \sim P_{\mathbf{x}_0}$ and $\boldsymbol{\xi} \sim P_{\boldsymbol{\xi}}$
to be known, while the dynamics parameterization $\boldsymbol{\nu} \sim P_{\boldsymbol{\nu}}$ may be unknown and require inference from  data. 
After solving \eqref{eq:pocp},
the resulting family of policy functions allows for direct adaptation to arbitrary initial states $\mathbf{x}_0$, problem parameterizations $\boldsymbol{\xi}$, and dynamics $\boldsymbol{\nu}$. 

\section{Zero-Shot Function Encoder-based Differentiable Predictive Control}

Function encoders learn compact representations of system dynamics in the form of unique latent coefficients that can be estimated online from data.
By conditioning DPC policies on these coefficients, our approach enables full differentiability while preserving computational efficiency. A diagram detailing our approach is shown in~\Cref{fig:FE-DPC_diagram}, and formalized in Algorithms~\ref{algo:FE-training}, and~\ref{algo:dpc-training}.

\vspace{0.5em}
\begin{minipage}[htb]{0.48\textwidth}
\begin{algorithm2e}[H]
\caption{Offline learning of FE basis functions for system dynamics}
\label{algo:FE-training}
\DontPrintSemicolon
\LinesNumbered
\KwIn{set of datasets $\mathcal{D}$ collected from  dynamics $\mathcal{F}$, learning rate $\alpha$}
Initialize neural basis functions $\mathbf{g}_1, \dots, \mathbf{g}_B$ with trainable parameters $\boldsymbol{\theta}_1, \dots, \boldsymbol{\theta}_B$ \\
\While{not converged}{
\For{$\mathcal{D}_l \in \mathcal{D}$}{
reset loss $L = 0$ \\
\For{$( \mathbf{x}_k, \mathbf{u}_k, \mathbf{x}_{k+1} ) \in \mathcal{D}_l$}{
$\mathbf{c} \gets (\mathbf{G} + \lambda \mathbf{I})^{-1} \mathbf{F} $\\
$ \mathbf{\hat{x}}_{k+1} \gets \mathbf{\hat{x}}_{k+1}$ from~\eqref{eq:FE NODE}  \\
$L \gets L + \| \mathbf{x}_{k+1} - \mathbf{\hat{x}}_{k+1} \|_2^2$ 
}
$\boldsymbol{\theta} \gets \boldsymbol{\theta} - \alpha \nabla_{\boldsymbol{\theta}} L$
}
}
\KwOut{trained basis functions $\mathbf{g}_1, \dots, \mathbf{g}_B$ }
\end{algorithm2e}
\end{minipage}%
\hfill
\hspace{0.4em}
\hfill
\begin{minipage}[htb]{0.46\textwidth}
\begin{algorithm2e}[H]
\caption{Offline learning  of parametric neural policies via DPC}
\label{algo:dpc-training}
\DontPrintSemicolon
\LinesNumbered
\KwIn{pre-trained NODE basis functions $\mathbf{g}_1, \dots, \mathbf{g}_B$, learning rate $\beta$}
Initialize policy network $\pi_{\mathbf{W}}$ with trainable parameters $\mathbf{W}$ \\
\While{not converged}{
\For{$\mathbf{x}_0 \sim P_{\mathbf{x}_0},\, \boldsymbol \xi \sim P_{\boldsymbol \xi}, \, \mathbf{c} \sim P_{\mathbf{c}}$}{
loss $L \gets 0$ \\
\For{$k = 0,1,N-1$}{
$\mathbf{u}_{k} \gets \pi_{\mathbf
W} (\mathbf{x}_k; \boldsymbol{\xi}, \mathbf{c})$ \\
$\mathbf{x}_{k+1} \gets \mathbf{x}_{k+1}$ from~\eqref{eq:FE-DPC:x} \\
}
$L \gets L$ from~\eqref{eq:FE-DPC:obj} \\
$\mathbf{W} \gets \mathbf{W} - \beta \nabla_{\mathbf{W}}L$
}
}
\KwOut{trained policy network $\pi_{\mathbf{W}}$}
\end{algorithm2e}
\end{minipage}

\subsection{Modeling the Dynamics Using Function Encoders}

We model the family of system dynamics using a function encoder (FE) that learns basis functions parameterized by neural ordinary differential equations (NODEs) as in \cite{ingebrand2024zero, ingebrand2025function}. This allows for efficient zero-shot online adaptation in the closed-form.

\paragraph{Offline learning of the FE basis functions.} 
We assume access to a set of datasets $\mathcal{D}$, consisting of historical trajectories of each $\mathbf{f} \in \mathcal{F}$ where $\mathcal{F}$ is the target Hilbert space. More precisely, we have $\mathcal{D} = \{ \mathcal{D}_1, \mathcal{D}_2, \dots, \mathcal{D}_r \}$, where each $\mathcal{D}_i$ corresponds to a dynamics function $\mathbf{f}(\cdot, \cdot; \boldsymbol{\nu}_i)$. Each dataset $\mathcal{D}_i$ consists of data in the form of $( \mathbf{x}_k^i, \mathbf{u}_k^i, \mathbf{x}_{k+1}^i )$. 
This enables data-driven approximation of different dynamics. 
The FE learns a set of NODE basis functions $\lbrace \mathbf{g}_{1}, \mathbf{g}_{2}, \ldots, \mathbf{g}_{B} \rbrace$ that span a subspace $\hat{\mathcal{F}} = \mathrm{span} \lbrace \mathbf{g}_{1}, \mathbf{g}_{2}, \ldots, \mathbf{g}_{B} \rbrace$ supported by the data. Given some $\boldsymbol{\nu}$, the dynamics are approximated as $\mathbf{f}  \approx\hat{\mathbf{f}} \in \hat{\mathcal{F}}$, which takes the form of a linear combination of the learned basis functions under time discretization,
\begin{equation}
    \mathbf{x}_{k+1} 
     = \mathbf{x}_k + \int_{t_k}^{t_{k+1}} \mathbf{f}(\mathbf{x}(t), \mathbf{u}(t); \boldsymbol{\nu}) \text{d} t  \quad
     \approx \quad \mathbf{x}_{k} + \int_{t_k}^{t_{k+1}} \sum_{j=1}^{B} c_{j}(\boldsymbol{\nu}) \mathbf{g}_{j}(\mathbf{x}(t), \mathbf{u}_{k}; \boldsymbol{\theta}_{j}) \text{d} t,\label{eq:FE NODE}
\end{equation}
where $\boldsymbol{\theta}_{j}$ are the network parameters of the basis network $\mathbf{g}_{j}$.
Importantly, the NODE basis functions $\lbrace \mathbf{g}_{1}, \mathbf{g}_{2}, \ldots, \mathbf{g}_{B} \rbrace$ do not depend explicitly on $\boldsymbol{\nu}$; as such, the dynamics function is uniquely determined by the coefficients $\mathbf{c} \in \mathbb{R}^{B}$. 
This learning procedure is formalized in Algorithm~\ref{algo:FE-training}. FE allow for flexible parameterization of the learnable neural basis functions and remain stable under different architecture choices, we refer to~\citep{ingebrand2025function} for detailed numerical studies. 

\paragraph{Online estimation of FE coefficients.} This can be done via regularized least squares optimization. The coefficients $\mathbf{c}$ to some $\mathbf{f}  \in \mathcal{F}$ can be computed in closed-form via the normal equation as $(\mathbf{G} + \lambda \mathbf{I})\mathbf{c} = \mathbf{F}$, where $\mathbf{G}_{ij} = \langle \mathbf{g}_{i}, \mathbf{g}_{j} \rangle$ and $\mathbf{F}_{i} = \langle \mathbf{f}, \mathbf{g}_{i} \rangle$ with $\langle \cdot, \cdot \rangle$ denoting the standard Euclidean inner product, can both be easily computed using Monte Carlo integration from a small amount of input-output data collected online. We apply Tikhonov regularization~\citep{golub1999tikhonov} in practice to ensure numerical stability. 
The weights $\{ \boldsymbol{\theta}_1, \dots, \boldsymbol{\theta}_B \}$ are frozen during online inference. 
This representation is key to our approach. By expressing the system in terms of coefficients $\mathbf{c}$, which we can efficiently compute online, we enable rapid adaptation of the dynamics, thereby making downstream control tasks viable.

\paragraph{Theoretical properties.} Function encoders provide a principled approximation for the dynamical model and are theoretically grounded. Prior work in \citet{ingebrand2025function} established that as the number of learned basis functions $B \to \infty$, the basis functions can span the entire Hilbert space supported by the data. Recent results in \citet{low2025function} strengthen this foundation by formalizing the Hilbert-space structure of the learned representation and by introducing finite-sample guarantees that bound the approximation error once the basis is fixed with an asymptotic rate
$\mathcal{O}(\frac{R}{\lambda \sqrt{m}})$, where $\lambda > 0$ is the regularization parameter; $m$ is the number of data points, and $R$ is a constant that depends on the number of basis functions $B$.
These guarantees let us represent new dynamics by estimating only their coefficients without retraining.

\subsection{Training the DPC Policy with Function Encoder Dynamics}

 Differentiable Predictive Control (DPC)~\citep{drgovna2022differentiable,drgovna2024learning} is a self-supervised learning method for offline policy optimization that is well suited for solving~\eqref{eq:pocp}. 
 DPC formulates the pOCP~\eqref{eq:pocp} as a differentiable program and trains a neural policy to approximate its closed-loop solution, amortizing the computational cost and eliminating the need for expensive online optimization, while preserving key properties of MPC, such as its model-based formulation, objective function, and constraint handling.
Having a known and differentiable parametric dynamical system is a key assumption in the DPC formulation. However, accurately approximating unknown dynamics while preserving the efficient online adaptability of DPC remains an open challenge that we aim to address in this work.
We achieve that by leveraging the differentiability and adaptability of FE-NODEs.

\paragraph{DPC with FE-NODE dynamics and neural policies.}
We combine FE-NODEs with DPC for offline learning of parametric control policies with online zero-shot generalization capability. 
We use a neural network $ \mathbf{u}_k = \pi_{\mathbf
W} (\mathbf{x}_k; \boldsymbol{\xi}, \mathbf{c}): \mathbb{R}^{n_x + n_\xi + B} \mapsto \mathbb{R}^{n_u}$ for policy representation with trainable parameters $\mathbf{W}$. Note that the unknown dynamics parameters $\boldsymbol{\nu}$ are replaced with a representation vector $\mathbf{c}$, consequently we obtain the following training problem in discrete-time form,
\begin{subequations}
\label{eq:FE-DPC}
\begin{align}
\min_{\mathbf{W}} \quad & \mathbb{E}_{\mathbf{x}_0 \sim P_{\mathbf{x}_0},\, \boldsymbol \xi \sim P_{\boldsymbol \xi}, \, \mathbf{c} \sim P_{\mathbf{c}} } \Bigg[
\sum_{k=0}^{N-1} \Big( \ell(\mathbf{x}_k, \mathbf{u}_k, \boldsymbol \xi) 
+ p(h(\mathbf{x}_k, \boldsymbol{\xi})) 
+ p(g(\mathbf{u}_k, \boldsymbol{\xi})) \Big) + p_N(\mathbf{x}_N)
\Bigg] \label{eq:FE-DPC:obj} \\
\text{s.t.} \quad 
& \mathbf{x}_{k+1} = \mathbf{x}_{k} + \int_{t_k}^{t_{k+1}} \sum_{j=1}^{B} c_{j} \mathbf{g}_{j}(\mathbf{x}(t),\pi_{\mathbf{W}}(\mathbf{x}_k; \boldsymbol \xi, \mathbf{c}); \boldsymbol{\theta}_{j}) \text{d} t. \label{eq:FE-DPC:x}
\end{align}
\end{subequations}
Here $p_N$ corresponds to $p_T$ in the continuous-time objective~\eqref{eq:pocp:objective}.
Constraints~\eqref{eq:pocp:g} are relaxed and incorporated as penalty terms in the objective function~\eqref{eq:FE-DPC:obj}. A common choice of penalty functions $p(\cdot)$ for equality constraints is 2-norm $\| \cdot \|_2^2$, while for inequality constraints it is $ \| \text{ReLU}(\cdot) \|_2^2$, where ReLU stands for rectifier linear unit function~\citep{glorot2011deep}. 
Weights can be used to balance terms in~\eqref{eq:FE-DPC:obj}; we omit them here for simplicity of notation.

\paragraph{Offline policy learning.}  DPC learns parametric control policy by minimizing~\eqref{eq:FE-DPC:obj} in a self-supervised manner. At each update, samples of $\mathbf{x}_0$ and $\boldsymbol{\xi}$ are drawn from their respective distributions, and the coefficient vector $\mathbf{c}$ are inferred from the collected dynamics data. To evaluate the objective function, full trajectory rollouts are required.
The FE-NODE dynamics allow for full differentiability of the computation graph, thereby enabling first-order, gradient-based methods for optimizing the policy network $\pi_{\mathbf{W}}$. In practice, automatic differentiation~\citep{baydin2018automatic} or the adjoint method~\citep{gholami2019anode,zhuang2020adaptive} can be used for gradient computation, depending on the problem specifics. We provide the detailed algorithm for DPC training in~\Cref{algo:dpc-training}.

\begin{wrapfigure}{r}{0.48\textwidth}
\vspace{-1em}
\begin{minipage}{\linewidth}
\setlength{\algomargin}{0pt}
\begin{algorithm2e}[H]
\DontPrintSemicolon
\SetAlgoLined
\KwIn{Trained bases $\{ \mathbf{g}_j \}_{j=1}^{B}$, $m$ online samples $\{ \mathbf{x}_k^l, \mathbf{u}_k^l, \mathbf{x}_{k+1}^l \}_{l=1}^{m}$ for  $\mathbf{f}$}
\tcc{Compute coefficients}
$\displaystyle\mathbf{G}_{ij} = \frac{1}{m} \sum_{l=1}^{m} \langle \mathbf{g}_i(\mathbf{x}_k^l, \mathbf{u}_k^l), \mathbf{g}_j(\mathbf{x}_k^l, \mathbf{u}_k^l) \rangle$  \\
$\displaystyle\mathbf{F}_{i} = \frac{1}{m} \sum_{l=1}^{m} \langle (\mathbf{x}_{k+1}^l-\mathbf{x}_{k}^l)/{\Delta t}, \mathbf{g}_i(\mathbf{x}_k^l, \mathbf{u}_k^l) \rangle$  \\
$\mathbf{c} \gets (\mathbf{G} + \lambda \mathbf{I})^{-1} \mathbf{F}$ \\
\tcc{Evaluate the policy}
$\mathbf{u}_{k} \gets \pi_{\mathbf{W}}(\mathbf{x}_k; \boldsymbol{\xi}, \mathbf{c})$  \\
\KwOut{Updated control inputs $\mathbf{u}_k$}
\caption{Online policy adaptation}
\label{algo:online-adapt}
\end{algorithm2e}
\end{minipage}
\vspace{-1em}
\end{wrapfigure}

\paragraph{Online policy adaptation.}  By conditioning the DPC policy on FE coefficients, a defining feature of our proposed approach is its zero-shot adaptivity in the policy space, even when the underlying dynamics are not fully known. Once the models are trained during the offline phase, no online model updates are required during evaluation. Only a limited number of observations are needed for dynamics inference and subsequent policy prediction. This makes our approach suitable for fast and accurate transfer of control policies to new and unseen problems. It is also well-suited to scenarios where online system switching may occur and real-time control predictions are required. We provide the outline for online adaptation in~\Cref{algo:online-adapt}, where we use $\mathbf{x}_{k+1} = \mathbf{x}_k + \Delta t \cdot \mathbf{f}(\mathbf{x_k, \mathbf{u}_k}; \boldsymbol{\nu})$ for simplicity. 

\section{Numerical experiments}
We conduct extensive numerical experimentation over four examples. These examples arise from various fields in control literature and range in dimensionality and complexity. We demonstrate the accuracy, efficiency, and robustness of our proposed approach.
In particular, we highlight the zero-shot generalization capability of our method through experiments involving online dynamics switching, showcasing its advantage over existing approaches.

\paragraph{Implementation details.}
For FE approximation, the required number of basis functions generally depends on the intrinsic dimensionality of the target function space. In practice, we aim to balance accuracy and speed.
Function encoder NODE models are configured on a per-example basis with the number of basis functions ranging between $11$ and $32$, each parameterized by multi-layer perceptrons (MLPs). 
For all examples, we use $100$ data points to compute the function encoder coefficients.
For the DPC policy network, a $4$-layer MLP with a hidden size of $256$ is used for all examples. We use the Adam optimizer~\citep{kingma2014adam} for all examples and train until convergence. We use the RK4 integrator for all systems, with the step size and other hyperparameters selected optimally for each example.
We implement MPC using CasADi~\citep{Andersson2019} using the true dynamics as baselines, referred to as the white-box (WB) models in the following sections to compare against our FE-DPC results.
All experiments were conducted on a single NVIDIA RTX 5090 GPU. The code is implemented in Pytorch~\citep{paszke2019pytorch} using the NeuroMANCER~\citep{Neuromancer2023} library. Hyperparameters are listed in~\Cref{sec:appendixhyper}.

\paragraph{Stabilizing a Van der Pol Oscillator.}
\begin{figure}[t]
    \centering
    \includegraphics[width=\linewidth,trim={0.15in 0pt 0in 0pt}, clip]{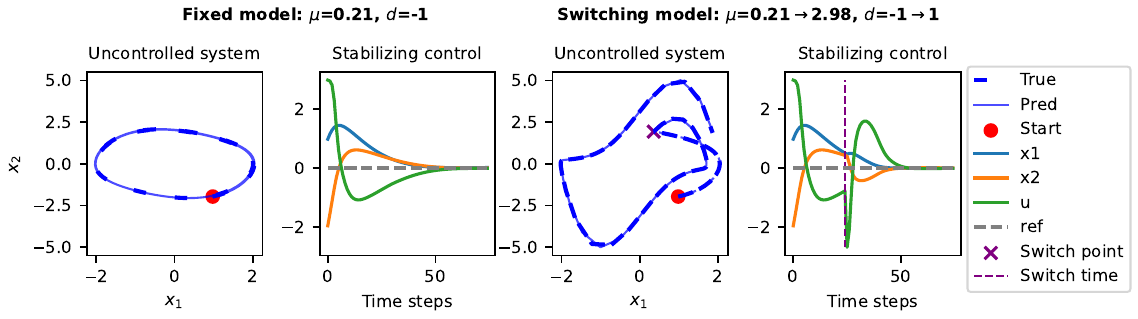}
    \vspace{-2em}
    \caption{Van der Pol oscillator dynamics under both controlled and uncontrolled scenarios using FE-DPC. \textbf{Left:} Uncontrolled and stabilizing results under a fixed dynamics setting. \textbf{Right: } Uncontrolled and stabilizing results with changing dynamics; in the example the dynamics switch after $25$ steps into the simulation. }
    \label{fig:VDP}
    \vspace{-1em}
\end{figure}
We first consider the task of stabilizing a nonlinear Van der Pol system, with dynamics
$\dot{x}_1 = d \cdot x_2$, $\dot{x}_2 = \mu(1-x_1^2)x_2 - x_1+u,$
where $[x_1, x_2]^\top \in [-2, 2] \times [-5, 5]$ is the state, $u \in [-3.0,\, 3.0]$ is the control, and parameters $\nu=(\mu,d)$ with $\mu\sim\mathcal{U}[0.1,3.0]$ and $d\in\{-1,1\}$ determine the dynamics.
The objective is to ``stabilize'' the system, that is,  $p_N(\mathbf{x}_N)=\|\mathbf{x}_N\|^2$ and $\ell(\mathbf{x}_k, \mathbf{u}_k, \boldsymbol \xi)=\| \mathbf{u}_k \|^2$ in eq.~\eqref{eq:FE-DPC:obj}.

Predictions from the learned function encoder and control policies from DPC are shown in~\Cref{fig:VDP}. The first two plots illustrate system behavior, without and with control, respectively, under fixed unknown dynamics that are identified in a zero-shot manner using least-squares estimation from measurement data. In the uncontrolled case, the predicted trajectories closely match the true trajectories simulated using the white-box model, validating the accuracy of the identified surrogate dynamics. The system is successfully stabilized within the prescribed input and state bounds when learned DPC policies are applied. The next two plots depict a scenario where the system parameters are switched during simulation. Despite an abrupt model switch, the proposed solution adapts online to stabilize the system. Importantly, this adaptation requires no retraining as the learned model is robust to parametric variations. 

\paragraph{Reference Tracking of a Two-tank System.}
\begin{figure}[t]
    \centering
    \includegraphics[width=0.9\linewidth, trim={0.1in 8pt 0.1in 7pt}, clip]{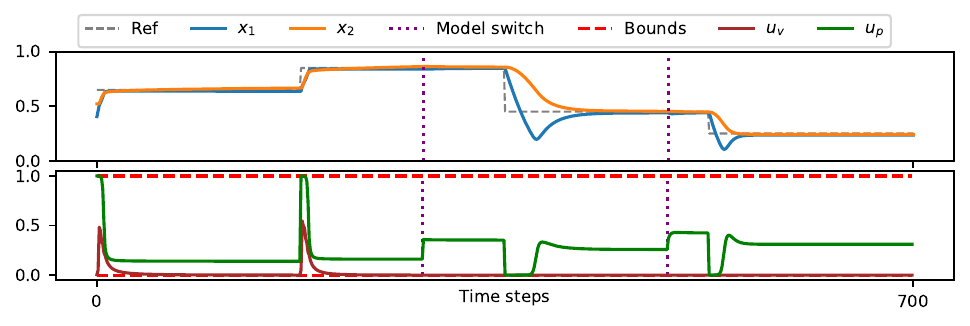}
    \vspace{-0.5em}
    \caption{Two-tank system reference tracking under multiple system switches using FE-DPC.}
    \vspace{-1em}
    \label{fig:TT}
\end{figure}
We consider level regulation of a two-tank system given the dynamics $\dot{x}_1 = d_1(1 - u_v)u_p - d_2\sqrt{x_1}$, $\dot{x}_2 = d_1u_vu_p + d_2\sqrt{x_1} - d_2\sqrt{x_2}$.
where $[x_1, x_2]^\top \in [0, 1]^2$ are the tank liquid levels, and control inputs $[u_p, u_v]^\top \in [0, 1]^2$ correspond to pump modulation and valve opening, respectively. The system is parameterized by inlet and outlet valve coefficients $d_1$ and $d_2$, i.e. $\nu=(d_1,d_2)$ where $d_1 \sim \mathcal{U}[0.06,\, 0.1]$ and $d_2 \sim \mathcal{U}[0.01,\, 0.06]$. The objective is to regulate the tank levels under different system parameterization to track desired reference values through coordinated modulation of the pump and valve control inputs. We have $p_N(\mathbf{x}_N)=\|\mathbf{x}_N-\mathbf{x}_{\text{ref}}(\boldsymbol{\xi})\|^2$ and $\ell(\mathbf{x}_k, \mathbf{u}_k, \boldsymbol \xi)=\|\mathbf{x}_k-\mathbf{x}_{\text{ref}}(\boldsymbol{\xi})\|^2+\|\mathbf{u}_k\|^2$ where $\mathbf{x}_{\text{ref}}(\boldsymbol{\xi})$ can vary.

\Cref{fig:TT} demonstrates accurate and stable reference tracking as both the system parameters and reference trajectories vary at inference time. The simulation spans $700$ time steps, during which the predicted and reference trajectories remain closely aligned, highlighting the accuracy of the learned dynamics model. The controller maintains consistent tracking performance through multiple system switches, showing strong robustness to changing conditions. Moreover, constraints on the control variable remain satisfied throughout the simulation.

\paragraph{Reference Tracking of a Glycolytic Oscillator (GO).}
\begin{figure}[b]
    \centering
    \includegraphics[width=0.9\linewidth, trim={0 8pt 5 12pt},clip]{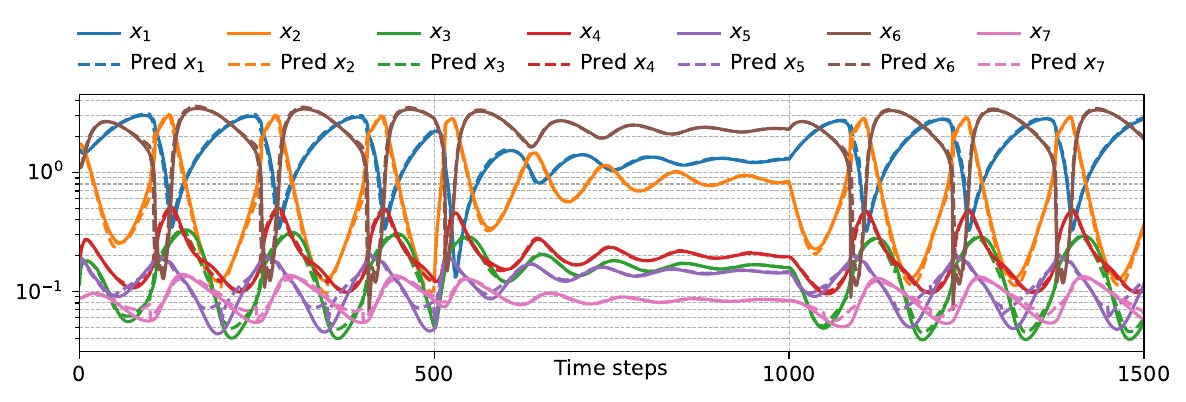}
    \vspace{-0.6em}
    \caption{True and predicted uncontrolled GO system dynamics. System parameterizations change every $500$ time steps, and predictions are calibrated against the true states every $50$ steps.
    }
    \label{fig:GO_autonomous}
    \vspace{-1em}
\end{figure}
We consider a highly nonlinear and stiff ODE system~\citep{daniels2015efficient}, which models the yeast glycolysis dynamics as:
\begin{equation}
\begin{cases}
\dot{x_1} &= J_0 - \frac{k_1 x_1 x_6}{1 + (x_6/K_1)^q} + u, \\
\dot{x_2} &= 2\frac{k_1 x_1 x_6}{1 + (x_6/K_1)^q} - k_2x_2(N - x_5) - k_6x_2x_5, \\
\dot{x_3} &= k_2x_2(N - x_5) - k_3x_3(A - x_6), \\
\dot{x_4} &= k_3x_3(A - x_6) - k_4x_4x_5 - \kappa(x_4 - x_7), \\
\dot{x_5} &= k_2x_2(N - x_5) - k_4x_4x_5 - k_6x_2x_5, \\
\dot{x_6} &= -2\frac{k_1 x_1 x_6}{1 + (x_6/K_1)^q} + 2k_3x_3(A - x_6) - k_5x_6, \\
\dot{x_7} &= \psi\kappa(x_4 - x_7) - kx_7.
\end{cases}
\label{eq:go_system}
\end{equation}
where
$[x_1, \dots, x_7] \sim 
\mathcal{U} (
[\begin{smallmatrix}
0.15, & 0.19, & 0.04, & 0.10 
0.08, & 0.14, & 0.05
\end{smallmatrix}],\;
[\begin{smallmatrix}
1.60, & 2.16, & 0.20, & 0.35 
0.30, & 2.67, & 0.10
\end{smallmatrix}] )$ represent concentrations of the seven biochemical species as states. We define control input $u \in [-4,\, 4]$.
For system parameterization, we set $J_0 = 2.5$, $k_2 = 6$, $k_3 = 16$, $k_4 = 100$, $k_5 = 1.28$, $k_6 = 12$, $q = 4$, $N = 1$, $A = 4$, $\kappa = 13$, $\psi = 0.1$, and $k = 1.8$ to be fixed, and $\nu=(k_1,K_1)$ where $k_1 \in \{ 80,90,100 \}$ and $K_1 \in \{ 0.5, 0.75 \}$ uniquely define each dynamics model. The system in Eq.~\eqref{eq:go_system} exhibits stiff behavior, requiring smaller integration steps and, consequently, a longer prediction horizon for control. Combined with its relatively high dimensionality, this makes the problem particularly challenging. Prediction horizon for FE-DPC and baseline WB-MPC is set to be $50$. In \Cref{fig:GO_autonomous}, the uncontrolled evolution of the dynamical system under different parameterizations of $k_1$ and $K_1$ is shown to alternate between stationary and oscillatory behavior. Evidently, the FE-NODE approximation achieves high accuracy under the prediction horizon. 
\begin{figure}[t]
    \centering
    \includegraphics[width=0.9\linewidth, trim={5pt 13pt 5 11pt},clip]{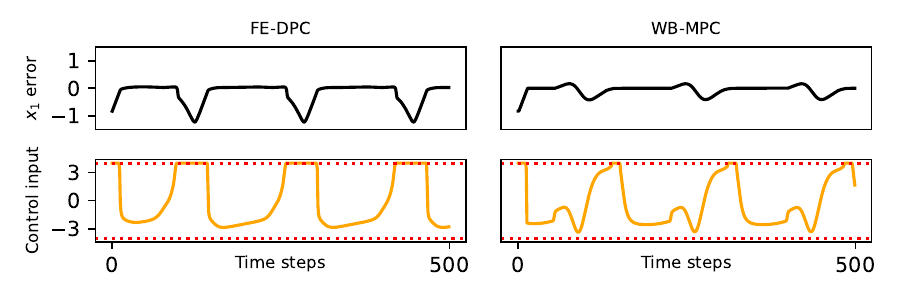}
    \vspace{-0.6em}
    \caption{FE-DPC (left) and WB-MPC (right) based reference tracking of $x_1$ state.}
    \label{fig:GO_qual}
    \vspace{-1em}
\end{figure}

For control, only the first state $x_1$ is directly actuated by the control input $u$, and it also serves as the reference-tracking state, while the remaining $6$ states evolve freely according to the nonlinear dynamics. The control objective is defined by $p_N(\mathbf{x}_N)=\|\mathbf{x}_N-\mathbf{x}_{\text{ref}}(\boldsymbol{\xi})\|^2$ and $\ell(\mathbf{x}_k, \mathbf{u}_k, \boldsymbol \xi)=\|\mathbf{x}_k-\mathbf{x}_{\text{ref}}(\boldsymbol{\xi})\|^2 +\|\mathbf{u}_k\|^2$ in eq.~\eqref{eq:FE-DPC:obj}. \Cref{fig:GO_qual} compares the performance of the controlled GO systems between the MPC solution under the known white-box system dynamics and the FE-DPC model prediction under online system identification.
In both cases, the control input saturates within bounds, and the controlled state exhibits an offset from the reference. WB-MPC solution achieves a lower tracking error since it optimizes under known system dynamics, while FE-DPC predicts a simpler plan that nonetheless accurately tracks the reference until the control saturates. As we show in Table~\ref{tab:controller_comparison}, computational efficiency is one of the main advantages of FE-DPC.

\paragraph{Controlling a Quadrotor.}
We consider a quadrotor flight control based on a $12$-dimensional nonlinear dynamic model~\citep{beard2008quadrotor, lopez2022arch}  that captures the vehicle's translational and rotational motion under thrust and torque inputs. The states include position, attitude, and velocity variables, while the $3$-dimensional control inputs correspond to the total thrust and roll/pitch torques. System parameters $\nu$ including mass and moments of inertia are randomly sampled and uniquely define the vehicle's motion. The control objective is to stabilize the vehicle at a target altitude of $0.4$m while having near-zero linear and angular velocities, we impose no running cost for the example. The relatively high dimensionality and nonlinearity of the dynamics make the problem particularly challenging. We include a detailed problem definition in~\Cref{sec:appendixquad}.

\begin{figure}[t]
    \centering
    \includegraphics[width=0.9\linewidth,trim={0.2in 3pt 0.1in 0pt}, clip]{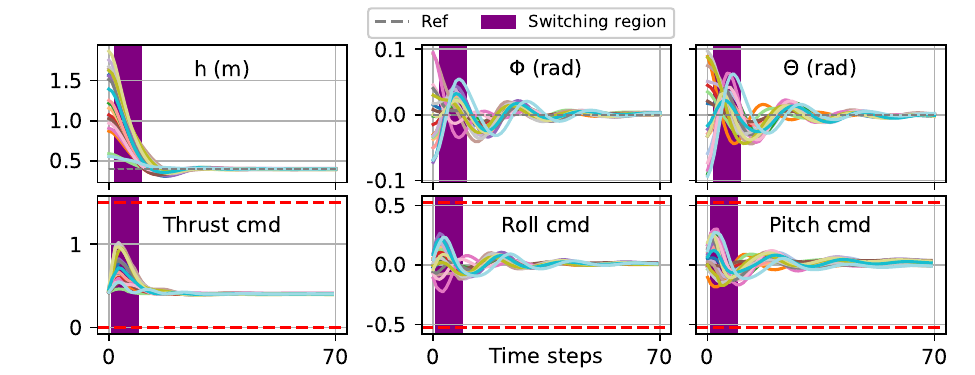}
    \vspace{-0.6em}
    \caption{$20$ quadrotor models with distinct dynamics parameterization are randomly initialized within state bounds. Each experiences a random dynamics switch between $2$-$20$ s. FE-DPC successfully stabilizes all models at the reference height of $h=0.4$m.}
    \label{fig:quad_qual}
    \vspace{-2em}
\end{figure}

Figure~\ref{fig:quad_qual} demonstrates the generalization capability of FE-DPC. The policy is evaluated on $20$ distinct system parameterizations, each initialized randomly within state bounds. To test robustness, each system undergoes an abrupt parameter switch at a random time between $2$ and $20$ seconds. In all cases, the controller maintains hover within the terminal cost tolerance $p_N(\mathbf{x}_N)$, indicating the learned parametric policies are robust to system changes under even complex dynamics.

\paragraph{Inference-time comparisons.}
One of the main advantages of our approach lies in its computational efficiency. 
Computation times measured during validation are presented in~\Cref{tab:controller_comparison}. Since baseline MPC cannot accommodate adaptive scenarios, we restrict this comparison to single parameter cases. MPC solvers can be slow due to the need for extensive online computation. At inference time, our method achieves a speed-up between $2.3$ and $71.5$ times over its MPC counterpart across the tested examples. The tracking error is evaluated using the mean squared loss (MSE) between predicted and reference states. While MPC attains the lowest tracking error due to its access to exact system dynamics, FE-DPC delivers comparable accuracy across all examples despite relying on learned system identification and control approximations.

\begin{table}[h]
\centering
\renewcommand{\arraystretch}{0.1}
\setlength{\tabcolsep}{5pt}      
\renewcommand{\arraystretch}{1.2}
\setlength{\tabcolsep}{6pt}
\begin{tabular}{llcccc}
\hline
\textbf{Algorithm} & \textbf{Metric} & \textbf{Van der Pol} & \textbf{Two Tank} & \textbf{GO} & \textbf{Quadrotor} \\
\hline
\multirow{2}{*}{FE-DPC} 
 & Error (MSE) & 0.002683 & 0.008452 & 0.180299 & 0.022003 \\
 & Time (s) & 0.53 & 1.13 & 5.89 & 1.93 \\ 
\hline
\multirow{2}{*}{WB-MPC} 
 & Error (MSE) & {0.002653} & {0.004164} & {0.032320} & {0.024208} \\
 & Time (s) & 1.21 & 6.75 & 136.07 & 155.85 \\ 
\hline
\end{tabular}
\vspace{-5pt} 
\caption{
Comparison of error (MSE) and computation time (s) for each benchmark.
}
\vspace{-2em} 
\label{tab:controller_comparison}
\end{table}

\section{Conclusion}
In this work, we present an end-to-end pipeline that combines function encoder-based system identification and differential predictive control to solve parametric optimal control problems. Our approach allows for zero-shot generalization to varying and unknown dynamics, given limited observations, and predicts corresponding policies. It is well-suited for problems requiring repeated online evaluation and enables real-time policy adaptation in response to system changes during inference.
We validate the efficiency and accuracy of our approach through extensive numerical experiments.
While the proposed framework demonstrates strong empirical performance across a range of nonlinear and complex control problems, some limitations remain.
Given the data-driven nature of the FE-NODE training, full-state information and sufficient data to represent the space of possible system dynamics is necessary for ensuring performance.
We also note that while we focus only on MLPs in this work, additional design choices such as~\citep{kim2021stiff,linot2023stabilized} may be explored to further improve performance; however, we will leave this for future work.

\acks{
Hassan Iqbal, Xingjian Li, and Krishna Kumar were partially funded by NSF 2339678 and 2321040.
J\'an Drgo\v na was supported by the Ralph O’Connor Sustainable Energy Institute at Johns Hopkins University. 
Tyler Ingebrand, Adam Thorpe, and Ufuk Topcu are supported in part by DARPA HR0011-24-9-0431, ONR N00014-25-1-2479, and NSF 2214939. Any opinions, findings, conclusions or recommendations expressed in this material are those of the authors and do not necessarily reflect the views of the funding organizations.
}

\bibliography{l4dc2026-sample}

\newpage
\appendix
\section{Supplementary Materials}
\label{sec:appendix}
\subsection{Controlling a quadrotor}
\label{sec:appendixquad}
In this section, we provide additional details on the Quadrotor control example, including the full dynamics model and corresponding control objectives. We use the quadrotor dynamics \citep{beard2008quadrotor, lopez2022arch} described by system of ODEs as,
\begin{equation*}
\left\{
\begin{aligned}
\dot{p}_n &= \cos\theta \cos\psi\,u 
           + (\sin\phi \sin\theta \cos\psi - \cos\phi \sin\psi)\,v 
           + (\cos\phi \sin\theta \cos\psi + \sin\phi \sin\psi)\,w, \\
\dot{p}_e &= \cos\theta \sin\psi\,u 
           + (\sin\phi \sin\theta \sin\psi + \cos\phi \cos\psi)\,v 
           + (\cos\phi \sin\theta \sin\psi - \sin\phi \cos\psi)\,w, \\
\dot{h}   &= \sin\theta\,u - \sin\phi \cos\theta\,v - \cos\phi \cos\theta\,w, \\
\dot{u} &= rv - qw - g\sin\theta, \\
\dot{v} &= pw - ru + g\cos\theta\sin\phi, \\
\dot{w} &= qu - pv + g\cos\theta\cos\phi - {F}/{m},\\
\dot{\phi}   &= p + \sin\phi\tan\theta\,q + \cos\phi\tan\theta\,r, \\
\dot{\theta} &= \cos\phi\,q - \sin\phi\,r, \\
\dot{\psi}   &= ({\sin\phi}/ {\cos\theta})\,q + ({\cos\phi}/ {\cos\theta})\,r, \\
\dot{p} &= ({(J_y - J_z)}/{J_x})\,q r + ({1}/{J_x})\tau_\phi, \\
\dot{q} &= ({(J_z - J_x)}/{J_y})\,p r + ({1}/{J_y})\tau_\theta, \\
\dot{r} &= ({(J_x - J_y)}/{J_z})\,p q + ({1}/{J_z})\tau_\psi. 
\end{aligned}
\right.
\label{eq:uav_dynamics}
\end{equation*}
where $p_n, p_e, h$ denote the inertial positions (north, east, altitude), 
$u, v, w$ are the body-frame linear velocities, 
$p, q, r$ are the body angular rates, 
and $\phi, \theta, \psi$ are the roll, pitch, and yaw Euler angles. 
The total thrust is $F$, and $\tau_\phi, \tau_\theta, \tau_\psi$ are the control torques about the body axes. Here, we make standard assumptions as in other works where $g=9.81$ and $\tau_\psi = 0$. The control inputs $\boldsymbol{\alpha}$ enter the system via, 
\begin{align*}
    & F = m g - 10(h - \alpha_1) + 3w,\\
    & \tau_\phi = -(\phi - \alpha_2) - p,\\
    & \tau_\theta = -(\theta - \alpha_3) - q.
\end{align*}
The control inputs are sampled as,

\begin{equation*}
[\alpha_1,\alpha_2,\alpha_3]^\top
\in [\alpha_{\min},\, \alpha_{\max}] 
= 
\begin{bmatrix}
0,\,-0.524,\,-0.524 \\
1.5,\,0.524,\,0.524
\end{bmatrix}^\top
\subset \mathbb{R}^3.
\end{equation*}
The system parameters are sampled as follows,
\begin{equation*}
m \sim \mathcal{U}[1.2,\,1.6], 
\quad J_x = J_y \sim \mathcal{U}[0.050,\,0.058], 
\quad J_z \sim \mathcal{U}[0.090,\,0.110].
\end{equation*}
The objective is for the quadrotor to maintain an altitude of $0.4$ with zero linear and angular velocities.
This setting provides a challenging benchmark due to the strong coupling and nonlinearity in the dynamics, making it a suitable test case for evaluating the proposed FE-DPC framework.

\subsection{Hyperparameters}
\label{sec:appendixhyper}
The training parameters and objective weights are presented as follows,
\begin{table*}[h]
\centering
\label{tab:hyperparameters}
\renewcommand{\arraystretch}{1.25}
\setlength{\tabcolsep}{6pt}

\begin{tabular}{p{2cm} p{3.5cm} p{3cm} p{1.2cm} p{5cm}}
\hline
\textbf{Problem} & \textbf{System parameters $\nu$} & \textbf{Parameter range / definition} & \textbf{\# Basis} & \textbf{Penalty / objective weights} \\
\hline

Van der Pol
& $\nu = (\mu, d)$
& $\mu \in [0.1, 3.0]$\newline
  $d \in \{-1,1\}$
& 11
& Control effort: $0.1$\newline
  State bounds: $10.0$\newline
  Terminal bounds: $20.0$ \\
\hline
Two-Tank
& $\nu = (c_1, c_2)$
& $c_1 \in [0.06, 0.10]$\newline
  $c_2 \in [0.01, 0.06]$
& 11
& State tracking: $5.0$\newline
  Control effort: $0.1$\newline
  State bounds: $10.0$\newline
  Terminal bounds: $10.0$ \\
\hline
GO
& $\nu = (k_1, K_1)$
& $k_1 \in [90, 100]$\newline
  $K_1 \in [0.5, 1.0]$
& 21
& State tracking (1st state): $2.0$\newline
  Terminal bounds: $200.0$ \\
\hline
Quadrotor
& $\nu = (m, J_x, J_y, J_z)$
& $m \in [1.2, 1.6]$\newline
  $J_x,J_y \in [.05, .058]$\newline
  $J_z \in [0.090, 0.110]$
& 32
& Position tracking: $50.0$\newline
  Velocity tracking: $200.0$\newline
  Euler-angle tracking: $100.0$\newline
  Angular-rate tracking: $150.0$\newline
  Control effort: $0.1$\newline
  State bounds: $10.0$\newline
  Control bounds: $10.0$\newline
  Altitude constraint: $100.0$ \\
\hline

\end{tabular}
\caption{Hyperparameters and penalty weights for the four benchmark control problems.}
\end{table*}

\end{document}